\documentclass[aps,showpacs,notitlepage,twoside,superscriptaddress,nofootinbib,amsmath,amssymb]{revtex4-1}  
\usepackage[textsize=tiny,color=gray!20]{todonotes}
\usepackage{graphicx}  
\usepackage{subfigure}
\usepackage{array}
\usepackage{longtable}
\usepackage{hyperref}
\usepackage{url}
\hypersetup{colorlinks,breaklinks,
citecolor=[rgb]{0.0,0.5,0.5},
    urlcolor=[rgb]{0.0,0.5,0.5},
    linkcolor=[rgb]{0.0,0.5,0.5}}
\DeclareMathOperator{\Tr}{Tr}
\DeclareMathOperator{\tr}{tr}
\def\ket #1{\vert #1\rangle}
\def\bra #1{\langle #1\vert}
\newcommand{\ketbra}[2]{\ensuremath{\ket{#1}\!\bra{#2}}}
\newcommand{\braket}[2]{\ensuremath{\bra{#1}  #2 \rangle}}

\newcommand{\Fq}{\ensuremath{\mathbb{F}_q}}

\newcommand{\Sigmamax}[1]{\langle \Sigma_{\textnormal{tot}}}

\newcommand{\Jam}{Jamio\l kowski }

\newtheorem{theorem}{Theorem}[section]
\newtheorem{lemma}[theorem]{Lemma}

\newtheorem{corollary}[theorem]{Corollary}
\newtheorem{definition}[theorem]{Definition}

\begin{document}

\title{Classical codes in quantum state space}
\author{Mark Howard}
\affiliation{%
Institute for Quantum Computing and Department of Applied Mathematics,
University of Waterloo, Waterloo, Ontario, Canada, N2L 3G1
}

\begin{abstract}
We present a construction of Hermitian operators and quantum states labelled by strings from a finite field. The distance between these operators or states is then simply related (typically, proportional) to the Hamming distance between their corresponding strings. This allows a straightforward application of classical coding theory to find arrangements of operators or states with a given distance distribution. Using the simplex or extended Reed-Solomon code in our construction recovers the discrete Wigner function, which has important applications in quantum information theory.

\end{abstract}

\maketitle

\section{Overview}
\begin{figure}[h!]
\centering
\subfigure[]{
\includegraphics[scale=0.8]{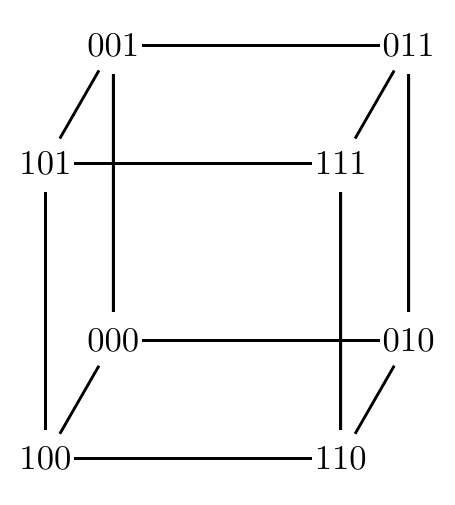}} \hspace{1cm}
\subfigure[]{
\includegraphics[scale=0.8]{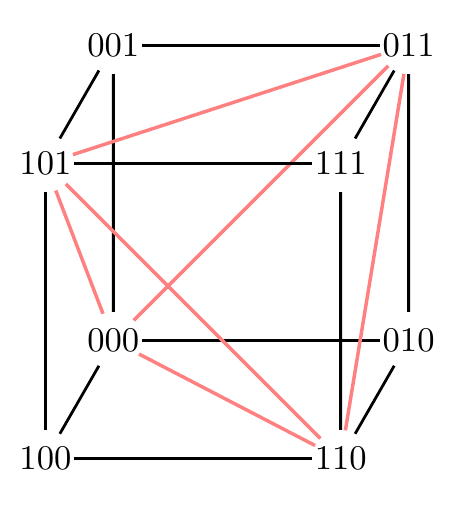}}\hspace{1cm}
\subfigure[]{
\includegraphics[trim=0cm 0.5cm 0cm 0cm, clip,scale=0.6]{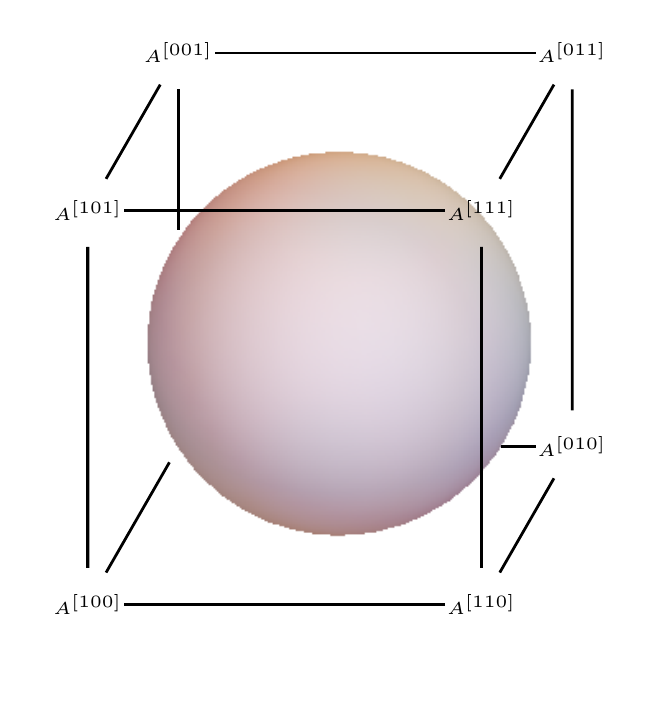}}
\caption{\label{fig:HammingCube} (a) Hamming cube for binary vectors of length 3, (b) the simplex code $\mathcal{C} \subset \mathbb{F}_2^3$ inscribed within it and (c) the facet operators $A^r$ corresponding to $r\in\mathbb{F}_2^3$ and their geometrical relationship to the Bloch ball (the subset of Hermitian operators corresponding to valid quantum states) for a single qubit.
}
\end{figure}

Figure \ref{fig:HammingCube}(a) depicts the binary Hamming cube -- all binary strings of length 3 where strings that differ by one element are one edge length apart, strings differing by two elements are two edge lengths apart etc. The number of differing elements between two strings is the Hamming distance and finding useful arrangements of $q$-ary strings (with prescribed mutual Hamming distances) is the subject of classical coding theory.
Let $q=p^n$ denote an integer that is a prime power. We will present a construction that associates $q$-ary strings with (i) Hermitian operators in Hilbert space of dimension $\text{dim}(\mathcal{H})=q$, and (ii) pure states in $\mathcal{H}^{\otimes 2}$. We find a remarkably simple relationship between the Hamming distance of strings and the Hilbert-Schmidt or Fubini-Study distance of the corresponding operators or states, respectively.    Because of the array of powerful coding-theoretic tools at our disposal, our construction may be useful for finding arrangements of quantum states or operators that would otherwise not be apparent.

Figure \ref{fig:HammingCube}(b) depicts the so-called simplex code, which is one member of family of $q$-ary codes that is well-defined for all prime powers $q$. Applying our construction to the codewords of this code (i.e., the vertices of the inscribed simplex) we find a set of operators that correspond to the phase point operators of Wootters' discrete Wigner function. States that have non-negative quasi-probability representation in this Wigner
function correspond to states that are not too far from any of the operators associated with the codewords of the code (see Fig.~\ref{fig:Intersecting Simplices}). This perspective on the Wigner function and its relationship with quantum state space may prove enlightening.

\section{MUBs and face operators}
We will adopt the notation of quantum information theory so that the standard basis has elements $\ket{k}:=\mathbf{e}_k \in \mathbb{C}^q=\mathcal{H}$ and $\braket{\cdot}{\cdot}$ is the inner product on $\mathcal{H}$.
Given an orthonormal basis $\mathcal{B}=\{\ket{0},\ldots,\ket{q-1}\}$, a unit vector $\ket{v}$ is called unbiased if $| \braket{v}{k} |=\frac{1}{\sqrt{q}}$ for all $0\leq k \leq q-1$. We will focus on Hilbert spaces of prime power dimension $q$ where it is known that $q+1$ (the maximal possible number) mutually unbiased bases always exist. For non-prime-power dimensions the number of MUBs is lower bounded by the largest component in a prime decomposition of $\text{dim}(\mathcal{H})$, but this is typically much lower than $\text{dim}(\mathcal{H})+1$. In subsequent sections we will be interested in connections between MUBs and classical coding theory so we find it  convenient to label MUB vectors with elements of the finite field, $\Fq$, containing $\text{dim}(\mathcal{H})=q$ elements. In fact it is quite natural to use $\Fq$ since many MUB constructions already use finite fields \cite{Wootters:1989,Klappenecker:2004,Ivonovic:1981} so that e.g. Gauss sums can be used to prove the required overlap constraints. A complete set of MUBs has one more basis than the number of field elements so we label this basis with $\infty$.

\begin{definition}{\textbf{Mutually unbiased bases:}\label{def:MUBs}}
A complete set of MUBs in a Hilbert space of dimension $\text{dim}(\mathcal{H})=q$ is given by $q+1$ orthonormal bases $\{\mathcal{B}_\infty,\mathcal{B}_0,\mathcal{B}_1,\ldots \}=\mathop{\bigcup}_{B\in \{\infty,\Fq\}}\mathcal{B}_B$, where each basis comprises $\mathcal{B}_B=\{\ket{\psi_B^V},V\in\Fq\}$, and overlaps obey
\begin{align}
|\langle \psi_B^V \vert \psi_{B^\prime}^{V^\prime}\rangle|=\frac{1}{\sqrt{q}}(1-\delta_{B,B^\prime})+\delta_{B,B^\prime}\delta_{V,V^\prime}.\label{eqn:MUBolap}
\end{align}
\end{definition}
Our results do not depend on the specifics of the mutually unbiased bases that we use. All that matters is their defining characteristic i.e., the pairwise inner products encapsulated in Eq.~\eqref{eqn:MUBolap}. In that sense, it is unnecessary that MUB vectors be labeled by elements of $\Fq$ since any consistent labeling will do. We suggest that our choice is as convenient as any other and has the additional merit that the labeling is physically meaningful in at least one case, which we discuss in Sec.~\ref{sec:The Weyl-Heisenberg Group}. (In the context of quantum information this particular MUB construction is important because all basis vectors are eigenvectors of Pauli/Weyl-Heisenberg operators.) A good survey of different MUB constructions in power-of-prime dimensions is provided by Kantor \cite{Kantor:2012}. Our construction also works without modification if we use a (necessarily incomplete) set of MUBs in $\text{dim}(\mathcal{H})=p^2$ that exclusively uses entangled basis vectors \cite{WvDMH:2011b}. The association between Hermitian operators and $\Fq$-valued vectors is given by the following definitions, whose name derives from a geometrical interpretation described in Sec.~\ref{sec:The Weyl-Heisenberg Group}.

\begin{definition}{\textbf{Facet operators:}\label{def:FacetOps}}
Using a complete set of mutually unbiased bases $\{\ket{\psi_B^V},V\in\Fq,B\in\{\infty,\Fq\}\}$ as in Definition \ref{def:MUBs}, a facet operator indexed by a vector $r \in \Fq^{q+1}$ is defined as
\begin{align}
A^{r=[r_\infty,r_0,r_1,\ldots,r_{B},\ldots]}=\sum_{B\in \Fq, \infty}\ketbra{\psi_B^{r_B}}{\psi_B^{r_B}}-\mathbb{I}_q. 
\end{align}
\end{definition}
For example in $\text{dim}(\mathcal{H})=3$ a possible facet operator $A^r$ with $r=[0,1,2,0]$ corresponds to choosing the zeroth vector from the computational ($\mathcal{B}_\infty$) basis, the first vector in the $\mathcal{B}_0$ basis, the second vector in the $\mathcal{B}_1$ basis and the zeroth vector in the $\mathcal{B}_2$ basis. Dropping the requirement that we select a vector from every basis we arrive at the definition of a Face operator,
\begin{definition}{\textbf{Face operators:}\label{def:FaceOps}}
Using a subset, of cardinality $|r|$ $(1\leq|r|\leq q+1)$, of a complete set of mutually unbiased bases, a face operator indexed by a vector $r \in \Fq^{|r|}$, is defined as
\begin{align}
A^{r}=\sum_{\substack{B\subseteq \{\infty,\Fq\}\\ |\{B\}|=|r|}} \ketbra{\psi_B^{r_B}}{\psi_B^{r_B}}-\left(\frac{|r|- \sqrt{q^2-q|r|+|r|}}{q}\right)\mathbb{I}_q. \label{eqn:PPO} 
\end{align}
\end{definition}
For example, we could drop the $\mathcal{B}_\infty$ and $\mathcal{B}_1$ bases from the previous example, and then $A^r$ with $r=[r_0,r_2]=[2,0]$ corresponds to taking the second vector from $\mathcal{B}_0$ and the zeroth vector from $\mathcal{B}_2$. The definition for face operators completely subsumes the previous one since facet operators correspond to the special case $|r|=q+1$. Nevertheless we have given them separate definitions as facet operators are the most interesting, and the simplification of the identity coefficient is not immediately apparent.

\section{Finite fields and $q$-ary codes}

A field is a non-empty set $\mathbb{F}$ of elements with abelian addition and multiplication, satisfying the usual axioms e.g. distributivity. We denote as $\Fq$ the finite field of order $q=p^n$ where $p$ is a prime and $n\geq 1$ is an integer. The smallest number of times the unit element $1\in \Fq$ must be added to itself to produce $0$ is the characteristic of the field, which is $p$, and consequently any element $\beta\in\Fq$ satisfies $p\beta=0$. If $n=1$ and $q=p$ then $\Fq \cong \mathbb{Z}_p:=\{0,1,\ldots,p-1\}$ -- the integers modulo $p$. When $n>1$ it is necessary to extend $\mathbb{F}_p$ to $\Fq$ with the addition of extra elements but we will not discuss the details of how this achieved. It will be sufficient to note that the nonzero elements of $\Fq$ form a cyclic group of order $q-1$ and a primitive element denoted $\alpha$ generates this whole group -- $\Fq /\{0\}=\{\alpha,\alpha^2,\ldots,\alpha^{q-1}=1\}$.  The $1$-dimensional vector space $\mathbb{F}_{p^n}$ is also an $n$-dimensional vector space over $\mathbb{F}_p$. Let $\tr:\mathbb{F}_{q=p^n} \mapsto \mathbb{F}_p$ be the trace map
\begin{align}
\tr(\beta):=\sum_{k=0}^{n-1}\beta^{p^k}
\end{align}
then a standard result (useful in the context of Weyl Heisenberg operators later) is that for any $\gamma \in \Fq$
\begin{align}
\sum_{\beta \in \Fq} \omega^{tr(\beta \gamma )}= q\delta_{\gamma,0} \qquad \text{where }\omega:=exp(2\pi i/p). \label{eqn:CharSum}
\end{align}

A $q$-ary alphabet, that is a set of $q$ distinct symbols, is naturally identified with elements of the finite field $\Fq$. A word of length $N$, $w\in\Fq^N$, is a string of $N$ symbols from $\Fq$ and clearly there are $q^N$ distinct words of this fixed length. The most general definition of a $q$-ary code is as a subset $\mathcal{C} \subseteq \Fq^N$ and the elements of $\mathcal{C}$ are called codewords (a good reference for all coding-related material is \cite{Macwilliams:1978}). The Hamming distance $0\leq \Delta(v,w)\leq N$ between two words $v,w \in \Fq^N$ is the number of positions in which $v$ and $w$ disagree. The Hamming distance is a metric on $\Fq^N$ so that expressions like $\Delta(u,w)\leq \Delta(u,v)+\Delta(v,w)$ hold. Using the Hamming distance we can define a ball/sphere of radius $r$ around any word $w$ via $\{v\in\Fq^N| \Delta(v,w)\leq r\}$. Roughly speaking, good codes consist of codewords $\mathcal{C} \subset \Fq^N$ where each codeword is the center of relatively large Hamming sphere, and this set of Hamming spheres fill the whole space without intersecting one another. The minimum distance $d(\mathcal{C})$ of a code is given by $d(\mathcal{C})=\min\{\Delta(x,y)|x\neq y \in \mathcal{C}\}$, and this is related to the radius of the empty Hamming spheres around each codeword. Two codes $\mathcal{C}$ and $\mathcal{C}^\prime$ are equivalent if they are related by trivial operations like permuting symbols or positions of codewords in a consistent way.  Codes can be either linear or non-linear with the former typically being more amenable to analysis and simple encoding procedures. A linear code of length $N$ has $q^k$ codewords for some integer $k\geq 0$ and is denoted $[N,k,d]$, whereas a nonlinear code has $M$ codewords and is denoted $(N,M,d)$. From a purely combinatorial point of view, linear codes may be outperformed by nonlinear codes.

The Hamming bound says that a $q$-ary code of block length $N$ and distance $d$ has a cardinality $|\mathcal{C}|$ that is upper bounded by following expression
\begin{align}
\text{Hamming Bound:} \qquad |\mathcal{C}|\leq q^N/\sum_{i=0}^{[\frac{d-1}{2}]}\binom{N}{i}(q-1)^i,
\end{align}
and codes that saturate this bound are perfect e.g., the Hamming codes mentioned later. The Singleton bound says that a code $\mathcal{C}$ of block length $N$ and minimum distance $d$ over a $q$-ary alphabet obeys
\begin{align}
\text{Singleton Bound:} \qquad |\mathcal{C}|\leq q^{N-d+1},
\end{align}
and codes that saturate this are maximum distance separable \cite{Singleton:1964} (MDS) e.g., the simplex codes mentioned later. 

The standard notation for the number of codewords of Hamming weight $i$ from the all zero codeword is
\begin{align}
A_i=|\{w\in \Fq^{N} |\Delta(w,0)=i\}|. \label{eqn:WeightDist}
\end{align}
and it should not be confused with a face operator (the subscript and context should avoid this issue). The set $\{A_i|0\leq i \leq N\}$ is the weight distribution of the code and is calculable using powerful tools like weight enumerators. Clearly for an $(N,M,d)$ code $\sum_{i=0}^{N}A_i=M$.

A code defines a vector space if and only if it is a linear code. A linear code encoding $k$ units of information is described by a generator matrix $G:\Fq^k \mapsto \Fq^N $ e.g., the Simplex code depicted in Fig.~\ref{fig:HammingCube} has a generator matrix
\begin{align}
G_{\text{simplex}}&=\left[\begin{array}{ccc}
1 & 0 & 1  \\ 
0 & 1 & 1
\end{array} \right]=\left[\begin{array}{ccc}
  & g_1 &  \\ 
   & g_2 &   
\end{array} \right],\\
\Rightarrow\ \mathcal{C}_{\text{simplex}}&=\{ag_1+bg_2|a,b\in\mathbb{F}_2\}, \\
&=\{(0,0,0),(1,0,1),(0,1,1),(1,1,0)\}.\label{eqn:binsimplex}
\end{align}

The simplex code is well defined for all prime powers $q$ and for all lengths of the form $N=(q^m-1)/(q-1)$ with parameters $[N=(q^m-1)/(q-1),k=m,d=q^{m-1}]$. The maximum length of a code that we may use in our construction corresponds to $m=2$ and we will often refer to this code as \emph{the} simplex code. This simplex code sometimes goes by the name (doubly) extended Reed-Solomon code. In any event, our simplex code has generator matrix (recall that $\alpha$ is a primitive element of $\Fq$)
\begin{align}
G_{\text{simplex}}&=\left[\begin{array}{cccccc}
1 & 0 & \alpha & \alpha^2 & \cdots & \alpha^{q-1} \\ 
0 & 1 & 1 & 1& 1 & 1
\end{array} \right]=\left[\begin{array}{ccc}
  & g_1 &  \\ 
   & g_2 &   
\end{array} \right],\\
\Rightarrow\ \mathcal{C}_{\text{simplex}}&=\{ag_1+bg_2|a,b\in\mathbb{F}_q\}.
\end{align}
The simplex code saturates the Singleton bound for all $q$ but only saturates the Hamming bound for $q=3$ where the simplex code is equivalent to the Hamming code
\begin{align}
G_{\text{Hamming}}=\left[\begin{array}{cccc}
1 & 0 & 1 & 2 \\ 
0 & 1 & 1 & 1
\end{array} \right]\qquad(q=3).
\end{align}

The fact that this $q=3$ code is doubly optimal (both MDS and perfect) arises from the following fact: the simplex code is dual to the Hamming code for all $q$ but these codes coincide (the code is self-dual) for $q=3$. The Hamming construction describes a family of codes with parameters $[N=(q^m-1)/(q-1),k=N-m,d=3]$ so once again we consider $m=2$ to describe \emph{the} Hamming code for our purposes. This has a generator matrix with $k=q-1$ rows i.e.,
\begin{align}
G_{\text{Hamming}}=\left[\begin{array}{cccccc}
1 & 0 & 0& \ldots & -\alpha^{q-1} & -\alpha^{q-1} \\ 
0 & 1 & 0& \ldots & -\alpha^{q-1} & -\alpha^{q-2} \\ 
0 & 0 & 1& \ldots & \vdots & \vdots \\ 
\vdots & \vdots & \vdots & \ldots & -\alpha^{q-1} & -\alpha^2 \\ 
\vdots & \vdots & \vdots & \ldots & -\alpha^{q-1} & -\alpha 
\end{array} \right].\label{eqn:Hamming}
\end{align}

For any linear code $\mathcal{C}$ we can define an equivalent code $\mathcal{C}^\prime=\mathcal{C}+w$ by adding a constant offset vector $w$ to each codeword so that both codes have the same distance distribution. A standard coding technique, typically used for decoding, is to partition $\Fq^N$ into cosets of a linear code, where each coset is identified (non-uniquely) by a coset leader $w$. 
This decomposition is depicted as a standard or Slepian array as in Table \ref{tab:Slepian} where we have given an example using the binary simplex code of Fig.~\ref{fig:HammingCube}(b).

{    \renewcommand{\arraystretch}{1.5}
\begin{table}[h!]
\begin{tabular}{|>{\centering}p{2.5cm}|>{\centering}p{1cm}|>{\centering}p{1cm}|>{\centering}p{1cm}|}
\hline 
Coset leader $w$ & \multicolumn{3}{c|}{Remainder of $\mathcal{C}+w$} \tabularnewline
\hline 
(0,0,0) & (1,0,1) & (0,1,1) & (1,1,0) \tabularnewline
\hline 
(0,0,1) & (1,0,0) & (0,1,0) & (1,1,1) \tabularnewline
\hline 
\end{tabular}
 \caption{\label{tab:Slepian} Slepian array partitioning $\mathbb{F}_2^3$ into cosets of the binary simplex code \eqref{eqn:binsimplex}. The top row corresponds to the vertices of the tetrahedron in Fig.~\ref{fig:HammingCube}(b), whereas the second row consists of the same strings translated by $(0,0,1)$. Together the simplex code and its translate exhaust all $8$ points of the binary Hamming cube.}
\end{table}
}

\section{Distances in quantum state space}

If we start with an operator of the form 
\begin{align}
A^{r}=\sum_{B\in \mathcal{B}}\ketbra{\psi_B^{r_B}}{\psi_B^{r_B}}-K\mathbb{I}_q 
\end{align}
then a fairly straightforward counting argument shows that
\begin{align}
\Tr(A^{r})&=|r|-qK,\\
\Tr\left((A^{r})^2\right)&=\frac{(q-1+|r|)|r|}{q}-2|r|K+qK^2,\\
&=q \text{ when } K=\frac{|r|\pm \sqrt{q^2-q|r|+|r|}}{q},
\end{align}
where the last line explains the somewhat peculiar choice of identity coefficient that we adopted in Def \ref{def:FaceOps}. Observe that face operators are clearly Hermitian $A^r=(A^r)^\dag$ since each term in the sum is manifestly so. We will examine the geometrical relationship between these face operators and it is assumed that the same bases are used in the construction of two face operators $A^r$ and $A^s$. The Hilbert-Schmidt inner product between these operators has remarkably simple expression, which is arguably the key insight of this work:
\begin{lemma}{}
Let $A^r$ and $A^s$ be face operators of the form \eqref{eqn:PPO}, in a  Hilbert space of dimension $\text{dim}(\mathcal{H})=q$, then
\begin{align}
\Tr\left(A^r A^s\right)&=q-\Delta(r,s)
\end{align}
where $\Delta(r,s)$ denotes the Hamming distance (number of differing elements) between vectors $r,s \in \Fq^{|r|}$. \label{lem:deltalemma}
\end{lemma}
\textit{Proof} Insert the face operator definition from Eq.~\eqref{eqn:PPO} and use the definition of mutually unbiased bases i.e.,
\begin{align}
|\langle \psi_B^V \vert \psi_{B^\prime}^{V^\prime}\rangle|^2=\frac{1}{q}(1-\delta_{B,B^\prime})+\delta_{B,B^\prime}\delta_{V,V^\prime}
\end{align}
along with the fact that $\sum_{j=1}^{|r|}\delta_{r_j,s_j}=|r|-\Delta(r,s)$.\footnote{If we want our face operators to have unit trace we can solve for a more general form
\begin{align}
A^{r}&=J\sum_{B\in \mathcal{B}}\ketbra{\psi_B^{r_B}}{\psi_B^{r_B}}-K\mathbb{I}_q \\
\Tr(A^{r})&=J|r|-qK=1\\
\Tr\left((A^{r})^2\right)&=\frac{J^2(q-1+|r|)|r|}{q}-2|r|JK+qK^2=q
\end{align}
so that
\begin{align}
J=\sqrt{\frac{q+1}{|r|}},\quad K=\frac{-1+\sqrt{|r|(q+1)}}{q}.
\end{align}
In that case we find
\begin{align}
\Tr\left(A^r A^s\right)&=q-\frac{q+1}{|r|}\Delta(r,s)
\end{align}
and the Hilbert-Schmidt and Fubini-Study distance measures can be derived from this.}

Since the operators $A^r$ are elements of the space of bounded linear operators, then the distance between two such operators can be characterized by the Hilbert-Schmidt metric.
\begin{corollary}{}\label{lem:HSdistance}
The Hilbert-Schmidt distance between two face operators $A^r$ and $A^s$ of the form \eqref{eqn:PPO} is
\begin{align}
D_{\textsc{HS}}\left(A^r, A^s\right)&:=\sqrt{\Tr[(A^r-A^s)^\dag(A^r-A^s)]}=\sqrt{2\left[q-\Tr\left(A^r A^s\right)\right]}=\sqrt{2\Delta(r,s)}
\end{align} 
\end{corollary}
We can also identify normalized pure quantum states with vectors $r\in\Fq^{|r|}$ by using the \Jam isomorphism \cite{Jam:1972,Choi:1975} and the distance between quantum states is once again simply related to the Hamming distance,
\begin{corollary}{}\label{thm:StateDist}
Let $A^r$ and $A^s$ be face operators of the form \eqref{eqn:PPO}, in a  Hilbert space of dimension $\text{dim}(\mathcal{H})=q$, then pure states $\ket{J^r}\in\mathbb{C}^{q^2}$ given by $\ket{J^r}=\left(\mathbb{I}\otimes A^r\right)\sum_{k\in\Fq} \ket{kk}/\sqrt{q}$ have trace distance and Fubini-Study distance
\begin{align}
D_{\textsc{TR}}\left(\ket{J^r},\ket{J^s}\right)&:=\sqrt{1-|\braket{J^r}{J^s}|^2}=\frac{1}{q}\sqrt{2q\Delta(r,s)-\Delta^2(r,s)}\ ,\\
D_{\textsc{FS}}\left(\ket{J^r},\ket{J^s}\right)&:=\sqrt{2-2|\braket{J^r}{J^s}|}=\sqrt{2(1-|1-\Delta(r,s)/q|)}\ ,
\end{align}
where the latter simplifies to $D_{\textsc{FS}}=\sqrt{2\Delta(r,s)/q}$ whenever $\Delta(r,s)\leq q$.
\end{corollary}
\textit{Proof} First note that, although face operators are not unitary in general, the \Jam isomorph obtained by applying $A$ to one half of a maximally entangled state produces a valid normalized pure state (which is not generally maximally entangled). This can be seen using $
\braket{J^r}{J^r}=\Tr((A^r)^2)/q=1$ and similarly
\begin{align}
\braket{J^r}{J^s}=\Tr(A^r A^s)/q=1-\Delta(r,s)/q.
\end{align}
The simplex code is equidistant with constant distance $\Delta=q$ between codewords so that $\{\ket{J^r},r\in\mathcal{C}_{\text{simplex}}\}$ forms a complete orthonormal basis in $\mathbb{C}^{q^2}$.

It is interesting to consider how evenly the set of states $\{\ket{J^r},r\in \Fq^{q+1}\}$ is distributed with resepect to the Haar measure. Finite sets of states approximating the uniform Haar measure are well studied and go by the name of state $t$-designs \cite{Ambainis:2007,Zauner:2011} (where $t\geq 1$ is an integer that quantifies how good the approximation is). The complete set of mutually unbiased bases described in Def.~\ref{def:MUBs} comprises a state 2-design. Numerical calculations suggest that the set $\{\ket{J^r},r\in \Fq^{q+1}\}$ provides a poor approximation to a Haar-uniform distribution of pure states in $\mathbb{C}^{q^2}$. For instance, the purity of the reduced state $\rho_1$ in a bipartite system quantifies how entangled the bipartite state is via $\frac{1}{q} \leq \Tr(\rho_1^2) \leq 1$ where the lower bound is saturated for maximally entangled states. A result due to Lubkin \cite{Lubkin:1978} states that a Haar-uniform distribution of bipartite pure states has average subsystem purity $\langle \Tr(\rho_1^2)\rangle_{\text{Haar}}=2q/(q^2+1)$, whereas we find
\begin{align}
q=3:\qquad&\langle \Tr(\rho_1^2)\rangle_{\mathbb{F}_3^{4}}=\frac{\left(\frac{3}{9}\right)9+\left(\frac{7}{9}\right)72}{3^4}=\frac{59}{81} ,
\end{align}
which suggests that entangled states may be under-represented in $\{\ket{J^r},r\in \Fq^{q+1}\}$.

The existence of a Hamming code \eqref{eqn:Hamming} with parameters $[q+1,q-1,3]$ means that for all prime power dimensions there exists a set of facet operators of size $|\{A\}|=q^{q-1}$ wherein any two elements obey 
\begin{align}
D_{HS}(A^r,A^s)&\geq \sqrt{6}.
\end{align}
For $q=2$ the Hamming code is simply $\mathcal{C}=\{(0,0,0),(1,1,1)\}$ and the facet operators correspond to opposite corners of a cube in the space of Hermitian operators as in Fig.~\ref{fig:Intersecting Simplices}. Using codewords of the Hamming code then the corresponding set of states obtained via Corollary \ref{thm:StateDist} obey
\begin{align}
D_{FS}(\ket{J^r},\ket{J^s})&\geq \sqrt{\frac{6}{q}}.
\end{align}
As well as knowing the minimum distance $d=3$ there exist powerful tools (e.g. weight enumerators \cite{Kim:2007}) for calculating the complete weight distribution \eqref{eqn:WeightDist} of codes such as this. In this way we can enumerate the number of states $\ket{J^s}$ at any given (discrete) distance from a particular reference state $\ket{J^r}$.

\section{MUBs and Facet Operators using the Weyl-Heisenberg Group}\label{sec:The Weyl-Heisenberg Group}

 The starting point for Weyl-Heisenberg operators in a Hilbert space of prime power dimension $\text{dim}(\mathcal{H})=q$ are the operators
\begin{align}
X(x)\ket{k}=\ket{k+x},\qquad Z(z)\ket{k}=\omega^{\tr(k z)}\ket{k}\qquad x,z,k\in\Fq,\ \omega:=exp(2\pi i/p), 
\end{align}
which compose as
\begin{align}
X(x)Z(z)X(x^\prime)Z(z^\prime)=\omega^{ \tr x^\prime z} X(x+x^\prime)Z(z+z^\prime).\label{eqn:composition}
\end{align}
A Weyl-Heisenberg (generalized Pauli) operator, indexed by $x,z\in\Fq$, is a product of these $X$ and $Z$ operators. From the composition law we observe that two Weyl-Heisenberg operators commute if and only if $\tr(xz^\prime-x^\prime z)=0$. The Weyl-Heisenberg operators generate a group that, modulo its center, has order $q^2$.
Consider a maximal abelian subgroup of this Weyl-Heisenberg group. Then any state that is a simultaneous eigenvector of all elements of this subgroup is a stabilizer state. 
Gross \cite{Gross:2006} showed that in a Hilbert space of dimension $q=p^n$ there are exactly $p^n\prod_{i=1}^n (p^i+1)$ distinct stabilizer states. Our MUB constructions below are comprised of basis vectors that are stabilizer states.

For odd prime powers, it turns out be convenient to impose a particular phase on the Weyl-Heisenberg operators so that they form the Weyl-Heisenberg group $\mathbf{D}$ of order $|\mathbf{D}|=q^2$,
\begin{align}
\mathbf{D}=&\{D_{x,z}:=\omega^{\tr \frac{xz}{2}}X(x)Z(z)|x,z,\in\Fq\} \qquad \left(\text{with }\ \frac{\beta}{2}=2^{-1}\beta\ \text{ where } 2^{-1}\in\Fq\right),
\end{align}
where individual group elements act as $D_{x,z}\ket{k}=\omega^{\tr \frac{xz}{2}+kz }\ket{k+x}$. 
Projectors onto rank-1 eigenstates of Weyl-Heisenberg operators (i.e., stabilizer states) can be constructed as \cite{Vourdas:2005,Gross:2007}
\begin{align}
\ketbra{\psi_B^V}{\psi_B^V}=\frac{1}{q}\sum_{k \in \mathbb{F}_q} \omega^{\tr(-k V)}D_{k,kB}, \label{eqn:StabBV}
\end{align}
so that
\begin{align}
D_{1,B}\ket{\psi_B^V}=\omega^{\tr(V)}\ket{\psi_B^V},\qquad V,B\in\Fq.
\end{align}
The set of states obtained by varying Eq.~\eqref{eqn:StabBV} over all $B,V \in \Fq$, along with the computational basis $\mathcal{B}_\infty=\{\ket{0},\ket{1},\ldots\}$ is a complete set of mutually unbiased bases.
One can check that the explicit form is given by
\begin{align}
\ket{\psi_B^V}=\frac{1}{\sqrt{q}}\sum_{k \in \mathbb{F}_q} \omega^{\tr(\frac{1}{2}Bk^2-Vk)}\ket{k}, \label{eqn:WHMUB}
\end{align}
and this is recognizable as the Ivanovic MUB construction \cite{Klappenecker:2004,Ivonovic:1981}.
Using the composition law Eq.~\eqref{eqn:composition} we can deduce
\begin{align}
D_{x,z}\ketbra{\psi_\infty^V}{\psi_\infty^V}D_{x,z}^\dag&=\ketbra{\psi_\infty^{V+x}}{\psi_\infty^{V+x}},\\
D_{x,z}\ketbra{\psi_B^V}{\psi_B^V}D_{x,z}^\dag&=\frac{1}{q}\sum_{k \in \mathbb{F}_q} \omega^{\tr(-k V)} D_{x,z} D_{k,kB} D_{x,z}^\dag,\\
&=\frac{1}{q}\sum_{k \in \mathbb{F}_q} \omega^{\tr(-k (V-z+xB))}D_{k,kB},\\
&=\ketbra{\psi_B^{V-z+xB}}{\psi_B^{V-z+xB}}.
\end{align}
Therefore the image of a facet operator under conjugation by a Weyl-Heisenberg operator is
\begin{align}
D_{x,z}A^{r}D_{x,z}^\dag&=A^{r+x[
1, 0, \alpha, \alpha^2, \ldots, \alpha^{q-1} 
]-z[
0, 1, 1,1, \ldots, 1
]}, \label{eqn:conjugate} \\
&=A^{r+xg_1-zg_2}\quad \text{ with } \left[\begin{array}{cccccc}
1 & 0 & \alpha & \alpha^2 & \cdots & \alpha^{q-1} \\ 
0 & 1 & 1 & 1& 1 & 1
\end{array} \right]=\left[\begin{array}{ccc}
  & g_1 &  \\ 
   & g_2 &   
\end{array} \right]
\end{align}
where $g_1$ and $g_2$ are the generators of the simplex code (a similar expression was already pointed out in the prime-dimensional case in \cite{Appleby:2008}). This is a very convenient way of understanding the orbit of facet operators under conjugation by the Weyl-Heisenberg group. It is also useful to have such a concise expression for the stabilizer states involved in the construction of a facet operator (for example, such a decomposition was used in \cite{Howard:2014} to construct a witness for quantum contextuality).

For even-prime-power dimension, i.e., $n$ qubits, it has been noted \cite{Wootters:1989,Klappenecker:2004,Godsil:2009} that an Ivanovic-type MUB construction \eqref{eqn:WHMUB} over $\mathbb{F}_{q=2^n}$ will not work without modification. Instead we must move to a slightly more general structure, the Galois ring $GR(4,n)$, which has $4^n$ elements and its associated Teichm\"uller set $\mathcal{T}=\{0,1,\xi,\xi^2,\ldots,\xi^{2^n-2}\}$ with $2^n$ elements. Each element $g \in GR(4,n)$ can be written $g=a+2b$ with $a,b\in\mathcal{T}$ and the trace map $\tr : GR(4,n) \mapsto \mathbb{Z}_4$ is defined via
\begin{align*}
\tr(g=a+2b)&=\sum_{k=0}^{n-1} a^{2^k}+2b^{2^k}.
\end{align*}
With these definitions we arrive at a MUB construction that appears formally very similar to the odd-prime-power case \eqref{eqn:WHMUB}
\begin{align}
\ket{\psi_B^V}&=\frac{1}{\sqrt{2^n}}\sum_{k \in \mathcal{T}} \omega_{4}^{\tr (Bk^2)+2\tr(Vk)}\ket{k}\qquad \omega_{4}:=exp(2\pi i/4)=i \label{eqn:TeichMUB}
\end{align}
except now our labels are elements of $\mathcal{T}$ rather than $\Fq$. For the purpose of investigating geometrical relationships between face operators, the distinction between $\mathcal{T}$-valued vectors and $\mathbb{F}_{2^n}$-valued vectors is irrelevant. From the form of the MUB vectors in Eq.~\eqref{eqn:TeichMUB} we can identify them as stabilizer states \cite{Dehaene:2003}, just as we had in the odd prime power case. For the even $q$ case we do not know of a similarly concise expression for the orbit of Weyl-Heisenberg operators acting on facet operators as we had in Eq.~\eqref{eqn:conjugate} although it should be possible. The Weyl-Heisenberg orbit of any $A^r$ with $r\in \mathcal{T}^{q+1}$ creates a simplex code e.g. for $q=4$ we have
\begin{align}
D_{x,z} A^{[0,0,\ldots,0]} D_{x,z}^\dag=\big\{A^r|\tr(r)\in
 &(0  0  0  0  0),
 (0  1  1  1  1),
 (0  2  2  2  2),
 (0  3  3  3  3),
 (1  0  1  2  3),
 (1  1  0  3  2),
 (1  2  3  0  1),
 (1  3  2  1  0),\nonumber \\
 &(2  0  2  3  1),
 (2  1  3  2  0),
 (2  2  0  1  3),
 (2  3  1  0  2),
 (3  0  3  1  2),
 (3  1  2  0  3),
 (3  2  1  3  0),
 (3  3  0  2  1)\big\},
\end{align}
where we are using coordinates $\tr(r)=(\tr(r_\infty),\tr(r_1),\ldots)\in{Z}^5_4$ rather than $r\in\mathcal{T}^5$.

\section{The discrete Wigner function}\label{sec:The DWF}

It is possible to represent finite-dimensional quantum states as probability distributions over a phase space of discrete points.  However, to recover all the predictions of quantum mechanics we must allow the probability distribution to take on negative values, in other words we must use quasi-probabilities. Such descriptions are referred to as quasi-probability representations, the most famous of which is the Wigner function. Wootters introduced a method of constructing discrete Wigner functions based on finite fields wherein vectors from a complete set of MUBs were put in one-to-one correspondence with the lines of the affine plane $AG(2,\Fq)$ \cite{Gibbons:2004}. The connection with our work is that Wigner function of state $\rho$ at some point in phase space is given by the expectation $\Tr(A\rho)$, where $A$ is one of the facet operators in Def.~\ref{def:FacetOps} and which go by the name phase point operators in the context of Wigner functions. The tools and terminology established in previous sections allow for an interesting interpretation of the relationship between phase point operators with each other and with quantum state space.

Wootters' discrete Wigner function (DWF) requires a set of $q^2$ trace-orthogonal phase point operators, which corresponds to a set of facet operators $\{A^r\}$ with pairwise Hamming distance $\Delta(r,s)=q$. We know that the codewords of $\mathcal{C}_{\text{simplex}}$ satisfy this constraint, as do the codewords of every translate $\mathcal{C}_{\text{simplex}}+w$ for fixed $w\in\Fq^{q+1}$. In this way we can obtain $q^{q-1}$ different DWF by partitioning $\Fq^{q+1}$ into $q^{q-1}$ cosets of the simplex code via a Slepian array as in Table \ref{tab:Slepian}. This partitioning is a coding-theoretic restatement of the concept of $q^{q-1}$ different ``quantum nets'' \cite{Gibbons:2004}. Hereafter, we will refer to a particular definition of DWF by its coset $w$, and it is understood that the MUBs used in constructing facet operators are those of Sec.~\ref{sec:The Weyl-Heisenberg Group}.

\begin{figure}[h!]
\centering
\subfigure[]{
\includegraphics[scale=0.25]{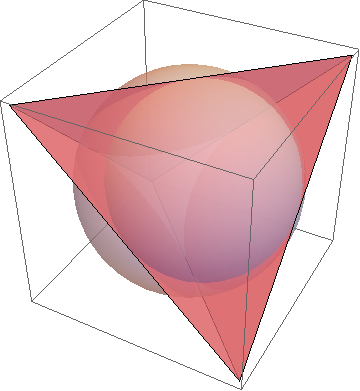}} 
\subfigure[]{
\includegraphics[scale=0.25]{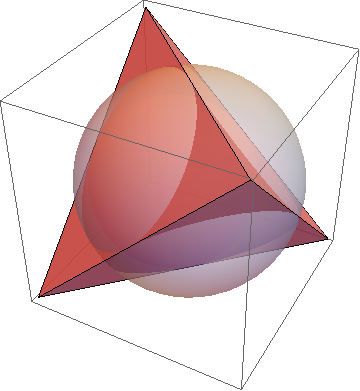}}
\subfigure[]{
\includegraphics[scale=0.25]{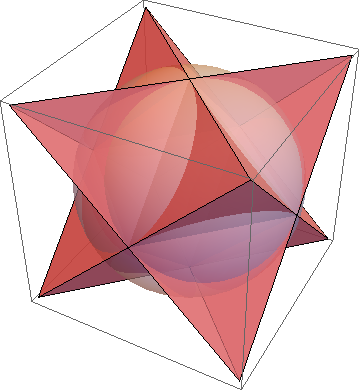}}
\subfigure[]{
\includegraphics[scale=0.25]{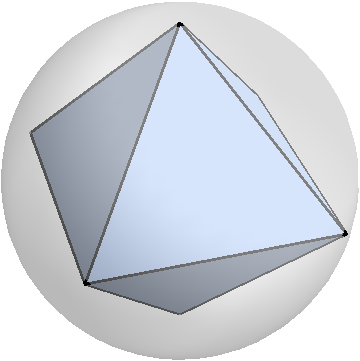}}
\caption{\label{fig:Intersecting Simplices}(a) Applying the facet operator construction to the simplex code $\mathcal{C}_{\text{simplex}}$ and (b) to a coset $\mathcal{C}_{\text{simplex}}+w$ (c) The intersection of these two simplices produces the stabilizer polytope (octahedron) (d) See also \cite{Galvao:2005,Appleby:2014,WvDMH:2011} for related geometrical discussions.}
\end{figure}

The Wigner function of $\rho$ at the point $(x,z)\in \Fq\times \Fq$ in phase space is denoted $W_{x,z}(\rho)$, and is defined via
\begin{align}
W_{x,z}(\rho)=\frac{1}{q}\Tr(A^{w+xg_1-zg_2} \rho) \qquad (\text{fixed }w \in \Fq^{q+1}) \label{eqn:DWF}
\end{align}
The quantity $\Tr(A^{w+xg_1-zg_2} \rho)$ is the Hilbert-Schmidt inner product between the operator $A$ and the density matrix $\rho$, and so demanding that $W(\rho)\geq 0$ is constraining $\rho$ to be close to $A$ in some sense. In fact the constraints $W_{x,z}(\rho)\geq 0\ \forall x,z \in \Fq$ describe a simplex in $\mathcal{H}$ with $q^2$ bounding facets (hence the name for facet operators).  In general our construction gives
\begin{align}
\text{ Simplex code in Hamming space}\quad \longleftrightarrow \quad \text{Simplex in Hilbert space}.
\end{align} 
Another geometrical object of interest is the single-particle ($q=p$) stabilizer polytope defined as the convex hull of all $p(p+1)$ stabilizer MUB vectors $\ket{\psi_B^V}$.  Cormick \emph{et al.} \cite{Cormick:2006} showed that a halfspace description of the single-qudit stabilizer polytope is given by 
\begin{align}
\text{Stabilizer polytope}:=\{\rho | \Tr(\rho A^r) \geq 0, \forall r \in \mathbb{F}_p^{p+1}\}. \label{eqn:StabPoly}
\end{align}
From the discussion in the previous paragraph we see that the stabilizer polytope is the intersection of all simplices associated with the simplex code and all its cosets (see Figure \ref{fig:Intersecting Simplices} for an illustration of the $p=2$ case).

 We are interested in non-negatively represented pure states, that is states $\ket{\phi}$ such that $W_{x,z}(\ketbra{\phi}{\phi})\geq 0\ \forall x,z \in \Fq$. In Figure \ref{fig:Intersecting Simplices}(a) we see that nonegatively represented pure qubit states in the DWF with $w=(0,0,0)$ are those that are both (i) on the surface of the Bloch sphere, and (ii) contained within the tetrahedron. Figure \ref{fig:Intersecting Simplices}(b) illustrates the same idea for the DWF with $w=(0,0,1)$.  Cormick \emph{et al.} \cite{Cormick:2006} showed that the only pure states that are non-negatively represented for all $q^{q-1}$ Wigner functions (simultaneously) are the $q(q+1)$ stabilizer MUB states $\ket{\psi_B^V}$ used in the DWF construction and this is illustrated in Figure \ref{fig:Intersecting Simplices}(c,d). When we move to qutrit (or any odd prime dimension) state space the story changes slightly from the qubit case we have depicted. There are now $q^{q-1}=9$ different DWF and we are guaranteed by \cite{Cormick:2006} that all $q(q+1)=12$ stabilizer states $\ket{\psi_B^V}$ are non-negatively represented no matter which DWF we use. However, a result by Gross \cite{Gross:2006} says that a pure state is non-negatively represented in the DWF with $w=\vec{0}$ if \emph{and only if} it is a stabilizer state. For qubits the set of non-negatively represented pure states is of finite measure, but for odd-prime qudits it is exactly the set of $p(p+1)$ stabilizer states. Moving on to multiple particles of odd prime dimension ($q$ is an odd prime power) then it seems that Gross' choice of DWF is the unique one obeying the discrete version of Hudson's Theorem \cite{Gross:2006}: a pure state is non-negatively represented if and only if it is a stabilizer state. A priori we know that at least $q(q+1)$ stabilizer states will be positively represented but this only represents an exponentially small (in $n$) fraction of all $p^n\Pi_{i=1}^n (p^i+1)$ stabilizer states. Hence the discrete Hudson theorem is a geometrically remarkable fact, as well as having practical relevance for questions surrounding fault-tolerant quantum computing \cite{Veitch:2012,Mari:2012}, resources theories \cite{Veitch:2014} and contextuality \cite{Howard:2014}. To see that $w=\vec{0}$ recovers Gross' choice of DWF insert \eqref{eqn:WHMUB} into $A^{r=[0,0,\ldots,0]}$ and simplify to obtain $\sum_{k\in \Fq} \ketbra{k}{-k}$ i.e., the discrete parity operator (the parity operator also forms the starting point for the continuous Wigner function \cite{Royer:1977}). This particular instance of Wootters' discrete Wigner function is also singled-out by its highly symmetric properties \cite{Gibbons:2004,Zhu:2015b,Chaturvedi:2010}.

  We hope that our way of analyzing these Wigner simplices and their relationships with each other and with the set of quantum states will prove enlightening. In principle, we could construct a Wigner-like representation using the Alltop \cite{Alltop:1980,Bengtsson:2014} MUB vectors 
$\ket{{}^{(a)\!  }\phi_B^V}=\frac{1}{\sqrt{q}}\sum_{k \in \mathbb{F}_q} \omega^{\tr(ak^3+\frac{1}{2}Bk^2-Vk)}\ket{k}$, which are equal to the Ivanovic MUB vectors for $a=0$ and unitarily equivalent but highly non-stabilizer \cite{Howard:2015,Andersson:2014} otherwise $(a\neq 0)$. One could also apply our Wigner simplex construction to unitarily-inequivalent MUB vectors \cite{Kantor:2012}. Note that Bengtsson and Ericsson \cite{Bengtsson:2005}, without restricting to stabilizer MUBs, have studied the equivalent of Wigner simplices and the stabilizer polytope (the complementarity polytope) and their relationship to quantum state space. Their construction of a simplex via mutually orthogonal Latin squares is isomorphic to our simplex code construction   \cite{Singleton:1964} in dimension $q=p^n$.

\section{Summary}

We identified a construction relating the Hamming distance between $q$-ary strings to the Hilbert Schmidt distance between certain Hermitian operators and the Fubini-Study distance between certain states. Any $q$-ary classical code of length up to $N\leq q+1$ is suitable and our hope is that the ability to use a vast array of coding-theoretic tools (distance distributions, automorphisms etc.) will prove useful in the quantum context. The types of operators and states that our construction provides are somewhat limited so it is unlikely that our results are directly applicable to outstanding open problems like the existence of symmetric informationally-complete positive operator-valued measures (SIC-POVMs). 

One topic for which our results are certainly relevant is the discrete Wigner function, where the expectation value of our operators are quasi-probabilities representing quantum states. The discrete Wigner function is both foundationally interesting as well as practically relevant for fault-tolerant quantum computing  \cite{Veitch:2012,Mari:2012,Veitch:2014}. We showed how a famous family of maximum distance separable codes, the simplex codes, when applied via our construction, reproduce the Wigner simplex in Hilbert space. More generally, our results represent a convenient tool for working with, and a novel way of thinking about,  Wootters' Wigner function in arbitrary prime-power dimension.

\section{Acknowledgements}

We thank Ingemar Bengtsson, Huangjun Zhu and Hammam Qassim for helpful comments on a previous draft.
We acknowledge financial support from the Government of Canada through NSERC
via the discovery grant program, as well as the U.~S.~Army Research Office
through grant W911NF-14-1-0103, and FQXI.

\end{document}